\documentclass[11pt]{article}
\usepackage{moriond,epsfig}

\def\MSbar {\hbox{$\overline{\hbox{\tiny MS}}\,$}}
\def\SDG{\hbox{\tiny SDG}}
\def\F{\hbox{\tiny F}}
\def\PT{\hbox{\tiny PT}}
\def\reg{\hbox{\tiny reg}}
\def\NP{\hbox{\tiny NP}}

\def\be{\begin{equation}}
\def\ee{\end{equation}}
\def\bea{\begin{eqnarray}}
\def\eea{\end{eqnarray}}

\begin{document}

\newcommand{\titlefootnote}{}
%
% FOR hep-ph version
% A
\begin{flushright}
  CERN-TH/2002-134\\
  June 2002
  \vspace*{-2.0cm}
\end{flushright}
\renewcommand{\titlefootnote}{\footnote{Talk presented at the XXXVII
    Rencontres de Moriond `QCD and high energy hadronic interactions',
    Les Arcs, France, and at the Workshop `Continuous Advances in QCD
    2002/Arkadyfest', Minnesota.}}

\vspace*{4cm}
\title{SOFT AND COLLINEAR RADIATION AND FACTORIZATION \\IN PERTURBATION THEORY AND BEYOND~\titlefootnote}

\author{ E. GARDI }

\address{TH Division, CERN, CH-1211 Geneva 23, Switzerland}

\maketitle\abstracts{
Power corrections to differential cross sections near a kinematic threshold are analysed by Dressed Gluon Exponentiation. Exploiting the factorization property of soft and collinear radiation, the dominant radiative corrections in the threshold region are resummed, yielding a renormalization-scale-invariant expression for the Sudakov exponent. 
The interplay between Sudakov logs and
renormalons is clarified, and the necessity to resum the latter whenever
power corrections are non-negligible is emphasized.
The presence of power-suppressed ambiguities in the exponentiation kernel suggests that power corrections exponentiate as well.
This leads to a non-perturbative factorization formula with non-trivial predictions
on the structure of power corrections, which can be contrasted with the OPE. 
Two examples are discussed. The first is event-shape distributions in the two-jet region,
where a wealth of precise data provides a strong motivation for the improved perturbative technique and an ideal situation to study hadronization. The second example is deep inelastic structure functions. In contrast to event shapes, structure functions have an OPE. However, since the OPE breaks down at large~$x$, it does not provide a practical framework for the parametrization of power corrections. Performing a detailed analysis of twist 4 it is shown precisely how the twist-2 renormalon ambiguity eventually cancels out. This analysis provides a physical picture which substantiates the non-perturbative factorization conjecture. 
}

\section{Dressed Gluon Exponentiation}

A classical application of QCD is the evaluation of semi-inclusive differential cross sections of hard processes depending on several scales.
We shall consider here cross sections that depend on a hard scale $Q$ and an intermediate scale $W$, both in the perturbative regime \hbox{$Q>W\gg~\Lambda$}.
Here $\Lambda$ represents the fundamental QCD scale. 
In~case of a large hierarchy, $Q\gg W$, there are large perturbative corrections containing logarithms, $\ln Q/W$. Typically, non-perturbative corrections are suppressed by powers of the {\em lower} scale~$W$. If the latter is not so large, such power corrections must be taken into account. From a theoretical point of view, power corrections are particularly interesting because of their relation to confinement~\cite{SVZ}.  

The first example is provided by event-shape distributions near the two-jet limit. Here the hard scale $Q$ is the centre-of-mass energy and the lower one is set by the shape variable, e.g. in the case of the thrust ($T$), it is $W^2=Q^2(1-T)^2$, the sum of the squared invariant masses of the two hemispheres. For $T\longrightarrow 1$ hadrons are produced in two narrow jets and large perturbative and non-perturbative corrections appear due to soft gluon radiation and hadronization.
The second example is structure functions in deep inelastic scattering (DIS) at  $x\longrightarrow 1$, where the hard scale is the momentum transfer $Q=\sqrt{-q^2}$ and the lower scale is the invariant mass of the hadronic system $W^2=(p+q)^2=Q^2(1-x)/x$, where $p$ and $q$ are the momenta of the proton and $\gamma^*$, respectively. For $x\longrightarrow 1$ the recoiling quark~\footnote{At large $x$ the gluon distribution is small.} develops into a narrow jet, and large corrections appear from the jet fragmentation process. 

\begin{figure}[htb]
  \begin{center} 
\vspace*{-10pt} \epsfig{height=12cm,width=8cm,angle=-90,file=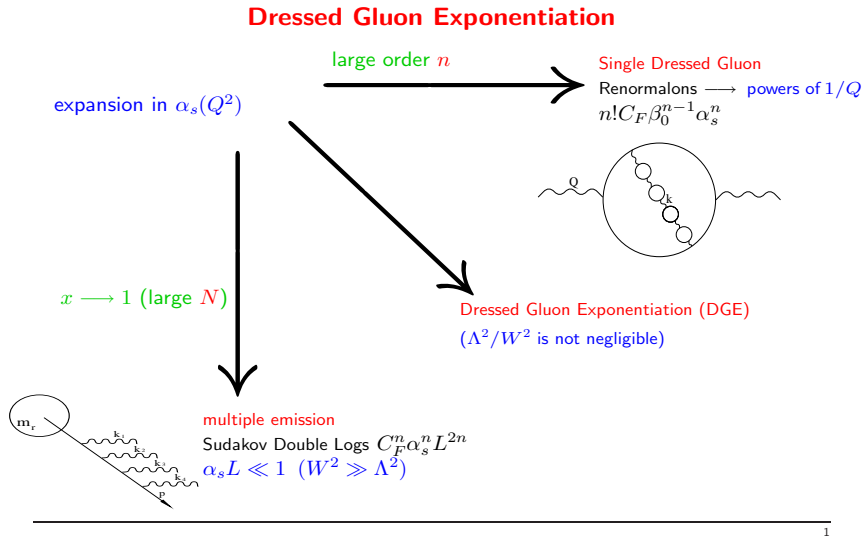}
\vspace*{-10pt}
  \end{center}
\caption{Different directions in summing the perturbative expansion in a two-scale problem.}
\label{PT}
\end{figure}

From the outset it is clear that resummation must be applied. Two relevant types of radiative corrections can be computed to all orders: renormalons and Sudakov logs. Renormalons appear from integration over the running coupling. Using the large $N_f$ limit, one calculates the diagrams where a {\em single} gluon is dressed by radiative corrections~\cite{Beneke}. These contributions dominate at large orders and exhibit the strongest sensitivity to infrared physics. Consequently they are useful in detecting power corrections~\cite{Beneke,DMW}. In the absence of an infrared cutoff, infrared renormalons make the perturbative expansion non-summable. The summation ambiguity is cancelled by non-perturbative corrections.
Sudakov logs, on the other hand, emerge from {\em multiple} emission of soft and collinear gluons. These contributions dominate the perturbative coefficients at large~$x$. At a fixed logarithmic accuracy, Sudakov logs can be summed to all orders~\cite{Contopanagos:1996nh,CTTW}, and, contrary to renormalons, they do not indicate non-perturbative corrections~\cite{Beneke:1995pq,Gardi:2001ny}. 
As shown schematically in fig.~\ref{PT}, these two classes of radiative corrections correspond to ``orthogonal'' sets of diagrams.
Both classes are relevant; however, none of the limits considered (large orders and large $x$, respectively) is appropriate in the threshold region, where both Sudakov logs and power corrections are important. 

The gap is closed by~\cite{Gardi:2001ny,DGE,Gardi:2002bg} Dressed Gluon Exponentiation (DGE). This is a resummation method that incorporates both types of diagrams by resumming renormalons in the Sudakov exponentiation kernel. Thus, contrary to the standard approach to Sudakov resummation the result is renormalization-scale-invariant. In the Sudakov exponent, renormalons appear through the enhancement of subleading logs. At any given power of the coupling $\alpha_s^n$, the leading log~\footnote{$L\equiv \ln N$, where $N$ is a moment index conjugate to $1-x$, so $N$ is large.},~$L^{n+1}$,  typically has a coefficient of order $1$, whereas the $L^k$ term (where $k\leq n$) 
 appears with a larger numerical factor~$\sim n!/k!$. Thus, the factorial growth in the exponent appears upon summing {\em all the logs}. In practice, subleading logs are not fully known and the state-of-the-art computation is restricted to a single dressed gluon (SDG) in the exponentiation kernel. In this case DGE reproduces the exact leading and next-to-leading logs (NLL)~\footnote{NLL require using the ``gluon bremsstrahlung'' effective charge~\cite{CMW} with a 2-loop renormalization-group equation.}, but only generates a certain class of subleading logs (next-to-next-to-leading logs and beyond) that can be regarded as an approximation to these coefficients. Not much is known about the accuracy of this approximation in general.
On the other hand, the factorial growth of subleading logs implies that a  resummation with a fixed logarithmic accuracy~\footnote{Such resummation is derived in the $N\longrightarrow \infty$ limit with $\alpha_s(Q^2) \longrightarrow 0$, so that $\alpha_s(Q^2)\cdot\ln N$ is small.} has a small range of validity. By construction it does not hold to power accuracy. If the latter is required, the additional resummation provided by DGE is necessary.

Analysing the large-order behaviour of the Sudakov exponent and the ambiguity associated with its resummation, one can access the dominant power corrections and obtain the information that is essential for their parametrization. The crucial difference in the way power corrections appear in this case, as compared with the standard OPE formulation, is that it is an overall factor multiplying~\cite{Korchemsky:1995is,Shape_function2,Dokshitzer:1997ew,Korchemsky:1999kt} the resummed perturbative result in moment (or Laplace) space, rather than an additive term. In DGE, such non-perturbative factorization is unavoidable, since the perturbative exponent by itself contains power suppressed ambiguities. The non-perturbative corrections, which compensate these ambiguities, must therefore exponentiate together with the perturbative logs. This exponentiation reflects the effect of multiple soft emission at the non-perturbative level. 
From the OPE point of view, the resulting non-perturbative factorization is highly non-trivial. It amounts to assuming that the dominant contribution at each twist
is proportional to the leading-twist matrix element. Moreover, the corresponding log-enhanced coefficient functions at higher twist must coincides with that of the leading twist. The matrix element information is essentially inaccessible by perturbative methods; however the higher-twist coefficient function can be computed, allowing one to check some of these far-reaching conclusions. A first step in this direction was recently taken~\cite{DIS} in the context of DIS structure functions.

\section{Event-shape distributions in the two-jet region}

A strong motivation for the improved resummation technique as well as for a systematic study of power corrections is provided by the very precise data on event-shape distributions in $e^+e^-$ annihilation. The goal is to have a handle on the parametrization of hadronization effects and to understand how they change depending on the observable considered. This will hopefully lead to better understanding of the hadronization process itself.

The DGE result for the single-jet mass distribution~\footnote{The single-jet mass distribution is used as an intermediate step in the evaluation of the thrust and the heavy-jet mass distributions. Note that this observable obtains additional non-global~\cite{Dasgupta:2001sh} corrections at the NLL level, which are absent in the thrust and the heavy-jet mass considered here.} is given by a Laplace integral,
\begin{eqnarray}
\label{dsig_drho}
\frac{1}{\sigma}\frac{d\sigma}{d\rho}(\rho,Q^2)=\int_{c}
\frac{d\nu}{2\pi i} \exp \left\{\rho\nu+ \ln J_{\nu}(Q^2)\right\},
\end{eqnarray}
where $c$ is an integration contour parallel to the imaginary axis. Here, terms that are not enhanced by logarithms of $\rho$ were discarded.  Eventually, such terms are included by matching~\cite{CTTW} the resummed expression to the fixed-order result, which is  currently available~\cite{EVENT2} numerically to next-to-leading order (NLO). The Sudakov exponent in~(\ref{dsig_drho}) is
\begin{eqnarray}
\label{lnJ}
\ln J_{\nu}(Q^2)\,=\, 
\int d\rho \left. \frac{1}{\sigma}\frac{d\sigma}{d\rho} (\rho,Q^2)\right\vert_{\SDG}
\left(  e^{-\nu \rho}-1 \right)
=\frac{C_F}{2\beta_0}
\int_0^{\infty} d{u}\,B_{\nu}({u}) \left(\frac{Q^2}{\bar{\Lambda}^2}\right)^{-u}\,
 \frac{\sin\pi{u}}{\pi{u}} \bar{A}_B({u}),
\end{eqnarray}
where $\bar{A}_B({u})$ depends only on the renormalization group equation for the coupling and  $\bar{\Lambda}$ corresponds to the ``gluon bremsstrahlung'' effective charge~\cite{CMW}. The Borel function is remarkably simple:
\begin{eqnarray}
\,B_{\nu}({u})
&=&\frac{2}{{u}}\left(\nu^{2u}-1\right)\Gamma(-2{u})-\left(\frac{2}{{u}}+\frac{1}{1-{u}}
+\frac{1}{2-{u}}\right)\left(\nu^{u}-1\right)\Gamma(-{u}).
\label{Borel_nu}
\end{eqnarray}
Using this distribution, and the assumption that the hemisphere masses are independent~\footnote{In the two-jet region, correlations between the hemispheres are suppressed perturbatively. It may play a more important role non-perturbatively~\cite{Korchemsky:2000kp,Belitsky:2001ij}. We neglect this effect~\cite{Gardi:2002bg}, still finding a good agreement with the data.}, both the thrust (we define $t\equiv 1-T$) and the heavy-jet mass ($\rho_H$) distributions are readily obtained,
\begin{eqnarray}
\label{t}
\frac{1}{\sigma}\frac{d\sigma}{dt}(t,Q^2)
&=&\frac{d}{dt}\int_c\frac{d\nu}{2\pi i
\nu}\,\exp\left\{\nu t+  2  \ln J_\nu(Q^2)\right\}\\
\label{rho_H}
\frac{1}{\sigma}\frac{d\sigma}{d\rho_H}(\rho_H,Q^2)
&=&\frac{d}{d\rho_H}
\left[\int_c\frac{d\nu}{2\pi i\nu}
\,\exp\left\{\nu\rho_H+ \ln J_\nu(Q^2)\right\}\right]^{ 2 }.
\end{eqnarray}

These resummation formulae suggest a specific way in which power corrections should be included~\cite{Gardi:2001ny,Gardi:2002bg}. First of all, since renormalon ambiguities appear in the exponent, $\ln J_{\nu}^{\PT}(Q^2)\, \longrightarrow \ln 
J_{\nu}^{\PT}(Q^2)\,+\, \ln J_{ \nu}^{\NP}(Q^2)$, power corrections appear as a factor in Laplace space, implying factorization and exponentiation of these terms, as previously suggested~\cite{Korchemsky:1995is,Shape_function2,Dokshitzer:1997ew,Korchemsky:1999kt}. Moreover, the particular structure of Borel ambiguities from the renormalon singularities in~(\ref{Borel_nu}) allows one to deduce the dependence of $\ln J_{ \nu}^{\NP}(Q^2)$ on $Q$ and $\nu$. There are two classes of corrections: (a) {\it odd powers} of $\bar{\Lambda}\nu/Q$ from the first term in~(\ref{Borel_nu}), which are related to large-angle soft emission; and (b) the {\it first two powers} of $\bar{\Lambda}^2\nu/Q^2$ from the second term in~(\ref{Borel_nu}), which are  
associated with collinear emission. The leading corrections are (a) and they can be resummed into a shape function of a single variable~\cite{Shape_function2}. The effect of the leading $\lambda_1\bar{\Lambda}\nu/Q$ power correction in the exponent is to shift the entire distribution~\cite{Shape_function2,Dokshitzer:1997ew}, whereas higher (odd) power corrections modify the shape of the perturbative spectrum.  

\begin{figure}[htb]
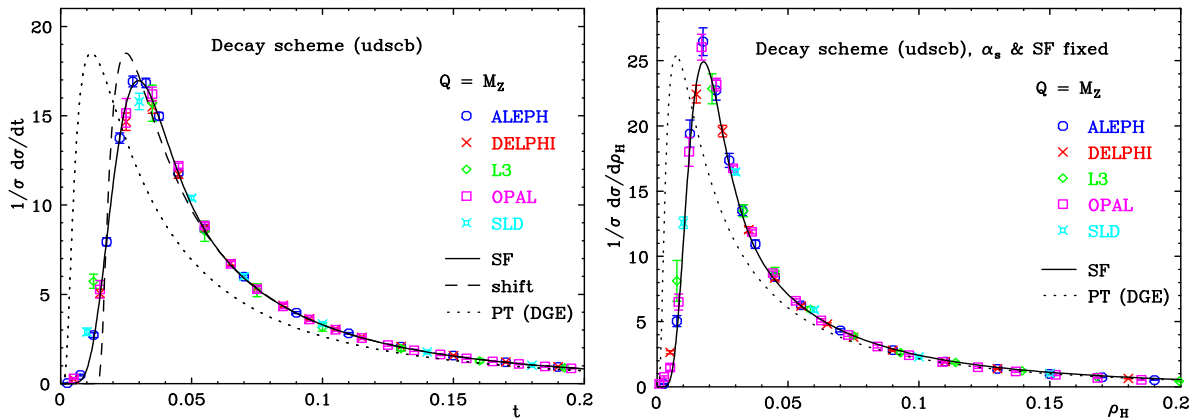

  \begin{center} \epsfig{width=5.5truecm,height=7.8truecm,angle=90,file=thrust_kcorr_mz.ps}   \epsfig{width=5.5truecm,height=7.8truecm,angle=90,file=rhoh_kcorr_mz_allfix.ps} 
\end{center}
\caption{Left: fit to the thrust distribution data at $Q=M_{\rm Z}$. The dotted line is the perturbative DGE result (principal-value regularization of the Borel sum), the dashed line shows a fit based on shifting the perturbative distribution (a single non-perturbative parameter) and the full line shows a shape-function-based fit. Right: the heavy-jet mass data compared with the predicted distribution based on the parameters fixed in the thrust fit. Here $\alpha_s^{\MSbar}(M_{\rm Z})=0.1086$.}
\label{t_rho}
\end{figure}

\begin{figure}[htb]
  \begin{center} 
\vspace*{-2pt} \epsfig{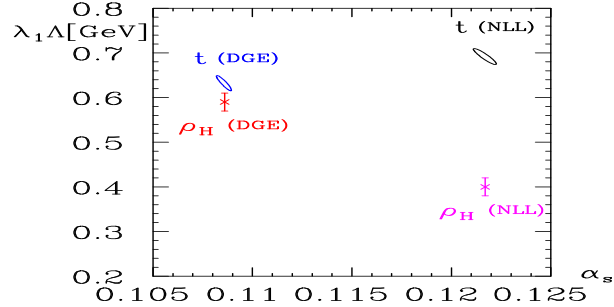}
\vspace*{-3pt}
  \end{center}
\caption{The leading non-perturbative correction on the scale $Q/\nu$, extracted from the $t$ and $\rho_H$ distributions.  Results based on DGE and on NLL resummation are 
shown. In each case $\alpha_s$ is fixed by the fit to the thrust.}
\label{correlation}
\end{figure}

Fitting the thrust distribution in a large range of $Q$ and $t$ values with such a shape function provides a strong test of the approach. Good fits were obtained to the world data~\cite{Gardi:2001ny,Gardi:2002bg,DGESHAPE}. An example~\cite{Gardi:2002bg} is shown in fig.~\ref{t_rho}. 
An even more stringent test is the comparison of the extracted parameters from the thrust and the heavy-jet mass distributions, both defined in the decay scheme~\cite{Salam:2001bd} in order to minimize the effect of hadron masses.  Assuming that~(\ref{t}) and~(\ref{rho_H}) hold non-perturbatively, the power corrections to the two distributions are associated with the same exponent and are therefore simply related. The agreement in the description of the two distributions is demonstrated in fig.~\ref{PT}, where the full line in the right frame shows not a fit but rather a calculated distribution for heavy-jet mass, where the parameters ($\alpha_s$ and the shape function) are fixed by the fit to the thrust.
A comparison of the leading power correction extracted from the two distributions is shown in fig.~\ref{correlation}. In addition to DGE, the figure shows a fit based on the NLL result. Contrary to DGE, in the NLL case there is no agreement between the two. This demonstrates the necessity to resum also the factorially enhanced subleading logs, as done by DGE, when a quantitative power correction analysis is done. Note also the significant impact on the extracted value of~$\alpha_s$. The best fit for the thrust distribution yields $\alpha_s^{\MSbar}(M_{\rm Z})=0.1086$. This value is consistent with that extracted from the average thrust upon performing renormalon resummation~\cite{Average_thrust}. 

\section{DIS structure functions at large Bjorken $x$}

\begin{figure}[htb]
  \begin{center}
\vspace*{-80pt}  
\epsfig{height=12cm,angle=90,file=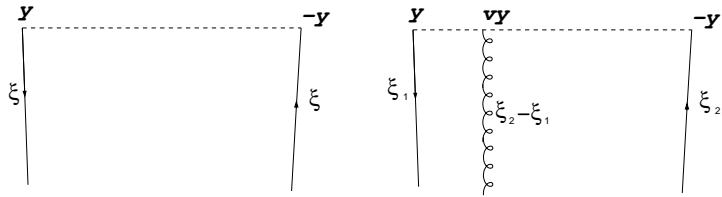} 
  \end{center}
\vspace*{-80pt}
\caption{Twist-2 and twist-4 light-cone operators whose hadronic matrix elements are the quark distribution and a correlation between quark--gluon and quark states, respectively. At twist 4 there are several other operators.}
\label{twist}
\end{figure}

In contrast to event-shape distributions, DIS structure functions have an OPE, by which power-suppressed contributions can be systematically identified and related to hadronic matrix elements~\cite{Jaffe:zw}.
The moments of the structure functions can be written as an expansion in inverse powers of the momentum transfer $q^2\equiv -Q^2$,
\begin{equation}
 \int_0^1 dx\,x^{N-1} F_2(x,Q^2) = C^{(2)}(N,\mu_{\F})\, \langle O^{(2)}(N)\rangle_{\mu_{\F}}
+\frac1{Q^2}\,\sum_{j}C^{(4)}_{j}(N,\mu_{\F})\, \langle O_j^{(4)}(N)\rangle_{\mu_{\F}}\,+\, \ldots,
\label{OPE}
\end{equation}
where $O^{(m)}_j$ are operators of twist $m$ with the appropriate quantum numbers. 
The OPE can be expressed in terms on non-local operators~\cite{JS,EFP,Jaffe,Balitsky:1989bk} defined on the light cone $y^2=0$ ($y$~is the Fourier conjugate of the $\gamma^*$ momentum $q$), as demonstrated in fig.~\ref{twist}.
Calculating the hadronic matrix elements of $O^{(m)}_j$ requires a full knowledge of the hadron structure. Thus, in practice these are simply parametrized and fixed by fit to experimental data. Moreover, most phenomenological analyses of structure functions are still restricted nowadays to the leading twist. On the other hand, the very existence of a non-perturbative definition puts the study of power corrections to structure functions on a firmer basis, as compared with event shapes. The OPE allows one to answer certain questions that are hard to address otherwise.
In particular, within the OPE one can trace the cancellation of infrared renormalon ambiguities~\cite{Beneke,UnPub}. This is not only important in principle, but, in fact, {\em essential} for any reliable measurement of higher-twist matrix elements, since the renormalon ambiguity from the resummation of $C^{(2)}(N,\mu_{\F})$ in~(\ref{OPE}) is of the same order as the next term in the expansion. It is well known that by studying the renormalization properties of these operators one can determine their large $Q^2$ scaling violation.
We shall also see that by calculating the higher-twist coefficient functions one can gain some insight into which partonic configurations may be relevant to certain kinematics. Both the anomalous dimension and the information encoded into the coefficient functions of the higher twist can be useful in identifying the dominant higher-twist contributions, without making too strong assumptions concerning the matrix elements themselves.  

\begin{figure}[htb]
  \begin{center}  
\vspace*{-40pt}
\epsfig{height=10cm,angle=-90,file=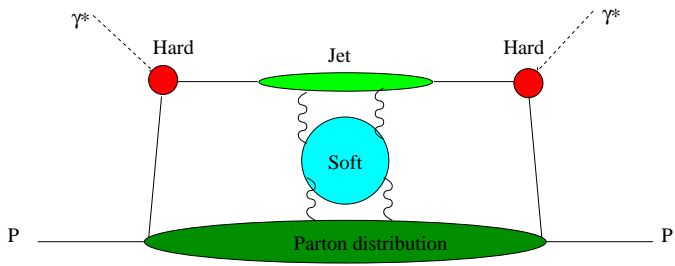}
\vspace*{-40pt}
  \end{center}
\caption{Factorization of DIS structure functions at large Bjorken $x$.}
\label{factorization}
\end{figure}

Let us now consider $F_2$ in the large-$x$ limit. In particular, consider the limit where $Q^2$ gets large but $W^2=Q^2(1-x)/x$ is held fixed at some low (yet perturbative) scale. At the level of the leading twist ($m=2$) the analysis of the structure function simplifies, since the gluon distribution can be neglected and only valance quarks contribute. The leading-twist factorization amounts to~\footnote{The way factorization is implemented in practice (through dimensional regularization) removes infrared singularities from the coefficient functions, making them well defined order by order. However, it is not equivalent to a rigid cutoff. Infrared effects do penetrate into the coefficient functions at orders as renormalons.} incorporating the effect of gluons softer than some scale $\mu_{\F}$ into the hadronic matrix element~$\langle O^{(2)}(N)\rangle_{\mu_{\F}}$, whereas the effect of gluons of higher virtualities go into the coefficient functions~$C^{(2)}(N,\mu_{\F})$. Independence of the structure function of $\mu_{\F}$ is guaranteed by an evolution equation stating that the logarithmic dependence of the renormalized operator on $\mu_{\F}$ is cancelled by that of the coefficient function. Moreover, radiative corrections can be  factorized into a hard subprocess, a jet subprocess, and a soft subprocess, which are mutually incoherent~\cite{Contopanagos:1996nh}, as shown in fig.~\ref{factorization}.
Interaction between the remnants of the target and the recoiling jet proceeds, at the level of the leading twist, only through the exchange of soft gluons (harder gluons contribute at higher twist), which cannot resolve the jet or the hadron structure.  

Since non-perturbative corrections at large $x$ go effectively in powers of $1/W^2$, the OPE
in~(\ref{OPE}) tends to break down and needs to be summed up. At first sight this seems an incredibly complex task, in particular, since the number of parameters involved increases sharply at large~$N$: at each twist ($m \geq 4$) the number of local matrix elements~\footnote{As $N$ increases, larger distances along the light cone become relevant. This makes the formulation of the OPE in terms of non-local matrix elements attractive. Indeed, using light-cone distributions, the analysis of twist~4 at large~$x$ becomes tractable.} grows as a power of $N$. 

Not only the need to compensate for the renormalon ambiguity at twist 2, but also the complexity of the twist expansion, and the difficulty to use it to parametrize power corrections, have led to renormalon-based phenomenology of structure functions. Renormalons resummation in the twist-2 coefficient function, at the level of a SDG (the large-$N_f$ limit), leads to the following ambiguity~\cite{DMW,Stein:1996wk,DasW_DIS,Maul:1997rz} to order $1/Q^2$,
\begin{eqnarray}
 \Delta\,\int_0^1dx\, x^{N-2}\,F_2 =  \frac{ C_F
}{2\beta_0}\left[4\psi(N+1)+4\gamma+\frac{2}{1+N}+\frac{12}{2+N}-8-N\right]
q_N\frac{\Lambda^2\delta}{Q^2},
\label{twist2_amb}
\end{eqnarray}
with $\delta\equiv \int_{\cal C}\!{du}/(1-u)$, where the contour ${\cal C}$ is the difference between different integration contours in the Borel plane, which avoid the singularity at $u=1$. 
Since $F_2$ is a physical quantity the ambiguity must cancel, and thus there must exist at twist 4 a contribution of the form~(\ref{twist2_amb}), which is proportional to the leading twist matrix element $q_N$. 
The renormalon-based model~\cite{DMW,Stein:1996wk,DasW_DIS,Maul:1997rz} amounts to assuming that this contribution dominates, and thus the entire twist 4 can be approximated by~(\ref{twist2_amb}), replacing the ambiguous  $\delta$ by a single fit parameter. No physical arguments have been given to support this dominance. From the OPE point of view it seems a very strong assumption: genuine multiparton correlations are absent here. One can also parametrize the twist-6 contribution based on the $1/Q^4$ renormalon ambiguity in the twist-2 coefficient function. At large $x$ this model does not apply: since we know that multiple emission is very important perturbatively, it is unlikely that power corrections will be entirely associated with a {\em single} dressed gluon. Instead, at large $x$ it is more appropriate to parametrize power corrections according to DGE, taking multiple gluon emission into account. But also in this case strong assumptions are made on the higher-twist contribution. 
To proceed we need to clarify the meaning of the renormalon dominance assumption, but we must also identify the dominant higher-twist contribution independently of any prejudice. Both these issues were recently addressed~\cite{DIS} at the level of twist 4.      

To clarify the meaning of the renormalon dominance assumption we must trace the cancellation of renormalon ambiguities within the OPE. The infrared renormalon ambiguity cancels against another ambiguity in the definition of higher-twist matrix elements, due to the mixing of the corresponding operators with the leading-twist operator~\cite{UV_dom,BBM,Beneke}.  
To make use of the OPE, one must regularize both sources of ambiguity in~(\ref{OPE}):
\begin{eqnarray}
C^{(2)}(N,\mu_{\F})&\longrightarrow& \left.C^{(2)}(N,\mu_{\F})\right\vert_{\reg}\,+
\, \frac{\Lambda^2\delta_{\reg}}{Q^2}\label{lt_amb}\\ 
\langle O_j^{(4)}(N)\rangle_{\mu_{\F}}&\longrightarrow&\left.\langle O_j^{(4)}(N)\rangle_{\mu_{\F}}\right\vert_{\reg}\,+\,\langle O^{(2)}(N)\rangle_{\mu_{\F}}\Lambda^2\delta_{\reg}^{(j)}.\label{ht_amb}
\end{eqnarray}
Here $\delta_{\reg}$ and $\delta_{\reg}^{(j)}$ in eqs.~(\ref{lt_amb}) and~(\ref{ht_amb}) represent the effect of changing the regularization prescription in defining the sum of the series in the twist-2 coefficient functions and the ultraviolet-divergent integrals in the renormalized twist-4 
operators, respectively. 
A consistent regularization guarantees the cancellation of all the $\delta$ terms leading to unambiguous predictions for the structure functions to power accuracy. It is clear from~(\ref{ht_amb}) that the mixing of twist~4 with twist~2 must be associated with {\em quadratic} divergence in the renormalization of~$O_j^{(4)}(N)$. Indeed, contracting the gluon in fig.~\ref{twist} to one of the quark lines, there appears a loop which is quadratically divergent in the ultraviolet, while the remaining operator is the \hbox{twist-2} one. An ambiguity emerges from the regularization of this loop. Summing the ambiguous ultraviolet contributions from all the relevant operators, each appearing with its own coefficient function, we recover an ambiguous expression, which is identical to~(\ref{twist2_amb}) but has an opposite overall sign~\footnote{The first example where this cancellation was demonstrated is the longitudinal structure function~\cite{Beneke,UnPub}. Recently~\cite{DIS} it was demonstrated in the case of $F_2$.}. This way, when the two are summed  in~(\ref{OPE}), a well-defined expression is obtained.

Note that while the renormalon ambiguity is associated with infrared scales, the presence of ambiguity in the higher twist is an ultraviolet property of the operator, which is not related to infrared physics. In conclusion, the renormalon ambiguity merely reflects the arbitrariness in separating contributions of different twists. The precise separation (or regularization of the renormalon sum) does not have any physical significance, and it should be regarded as complementary to the standard factorization used to define separately the coefficient functions and the operator matrix elements at each twist. 

From this discussion it follows that the renormalon dominance assumption should be interpreted, within the OPE, as the assumption that the ultraviolet 
divergent contribution~\cite{UV_dom,BBM,Beneke}, the one which mixes under renormalization with the leading twist, dominates the higher twist. 
Independently of this assumption we conclude that any treatment of higher twist which fails to deal with renormalon resummation (and the corresponding regularization of higher twist) is bound to be ambiguous. 

A priori, it is natural to expect that terms which mix with the leading twist will be of the same order of magnitude as other higher-twist effects. Ultraviolet dominance is the assumption that the former dominate. One can also imagine a scenario opposite to ultraviolet dominance, where the matrix elements are much larger than their ambiguous part. This point of view was adopted in the framework of QCD 
sum rules~\cite{SVZ}. The success of renormalon-based phenomenology in various applications calls for reconsideration of this assumption also in the framework of the sum rules. In general, similarly to DIS higher twist, condensates should be extracted when performing renormalon resummation; they can be assigned numerical values only within a given regularization prescription for the renormalons.

Let us now return to the large-$x$ limit and study the twist-4 contribution to $F_2$. Examining the coefficient functions of the corresponding light-cone operators at leading order we find~\cite{DIS} that a significant simplification occurs at large~$x$. This simplification is due to the fact that certain partonic configurations dominate the entire twist-4 contribution. In particular, we find that the dominating final state is that of a single energetic quark, just as at twist 2. The final state in which a quark and a gluon share the momentum is subdominant -- it is suppressed by a power of $1-x$. Thus the difference between the leading twist and twist~4 at large $x$ is restricted to  the initial states. However, also here a great simplification occurs. Representing the coefficient functions of the twist-4 operator in fig.~\ref{twist} in terms of the longitudinal momentum fractions of the quarks $\xi_1$ and $-\xi_2$, the dominant contribution arises from the region where the gluon momentum fraction is small $\xi_2-\xi_1\longrightarrow 0$, since the coefficient functions are singular at this point. Consequently the Heisenberg uncertainty principle implies that the quark--gluon--quark correlator     
$\langle p \vert O_j^{(4)}(v,y)\vert p \rangle_{\mu_{\F}}$ is essentially independent of the position of the gluon field on the light cone~$(v)$, and it effectively becomes a function of the light-cone separation between the quarks~($py$), just as the leading-twist matrix element $\langle p \vert O_j^{(2)}(y)\vert p \rangle_{\mu_{\F}}$. In conclusion, the configurations that dominate the twist-4 contribution to $F_2$ at large $x$, in both the final and initial states, make it twist-2-like. It is therefore natural to conjecture that ultraviolet dominance indeed holds, namely that the dominant ingredient in
twist~4 is the part that mixes with the leading twist. 

While justifying ultraviolet dominance at large~$x$, this picture does not apply to moderate or small values of $x$, and therefore it does not support the application of the renormalon model where eq.~(\ref{twist2_amb}) is assumed to represent the $N$ dependence of twist~4. 

The essence of the simplification we identified at twist~4 at large $x$ is that the dominant multiparton correlation measured by $F_2$ in such kinematics is still associated with the leading twist. This is most naturally realized through ultraviolet dominance. Gluons of momentum scale 
of order $W$ which are exchanged between the jet and the remnants of the target simply cannot resolve the full multiparton correlation function. Their interaction can be associated with ``power-like evolution'', similarly to the way the interaction of soft gluons at the leading twist is associated with logarithmic evolution. 

Assuming that ultraviolet dominance holds to all orders in the twist expansion,
a non-perturbative factorization formula (valid up to perturbative and non-perturbative corrections that are suppressed by $1/N$) emerges from the leading contributions to each twist:
\begin{eqnarray}
\!\!\!\!\int_0^1dx\, x^{N-1}{F_2}(x,Q^2)
&=&H\left({Q^2}\right)\,J\left({Q^2}/{N};{\mu_{\F}^2}\right)
\,\,q_N(\mu_{\F}^2) \left[1+\kappa_1\,\frac{N\Lambda^2}{Q^2} +\cdots\right] \\ \nonumber
&=&H\left({Q^2}\right)\,J\left({Q^2}/{N};\mu_{\F}^2\right)\,\,
q_N(\mu_{\F}^2)  \,J_{\NP}\left({N\bar{\Lambda}^2}/{Q^2}\right),
\end{eqnarray}
where $q_N$ is the twist-2 quark matrix element, $H$ and $J$ are the hard and jet components in the twist-2 coefficient function, and $\kappa_i$ are target-dependent non-perturbative parameters. In the second line all the non-perturbative corrections on the scale $Q^2/N$ are resummed into a shape function of a single argument, similarly to the parametrization of non-perturbative effects in event-shape distributions.  

DGE can now be applied to calculate $J$. Using the standard ${\overline{\rm MS}}$ factorization with $\mu_{\F}=Q$, we obtain
\begin{eqnarray}
\label{J}
J(Q^2/N,\mu_{\F}=Q)\,=\,\exp\left\{\frac{C_F}{2\beta_0}
\int_0^{\infty} d{u}\,B_{N}({u}) \left(\frac{Q^2}{\bar{\Lambda}^2}\right)^{-u}\,
 \frac{\sin\pi{u}}{\pi{u}} \bar{A}_B({u})\right\},
\end{eqnarray}
with~\cite{DGE}
\begin{eqnarray}
\,B_{N}({u})
&=&-\left(\frac{2}{{u}}+\frac{1}{1-{u}}
+\frac{1}{2-{u}}\right)\left(N^{u}-1\right)\Gamma(-{u})-\frac{2}{{u}}\ln(N),
\label{Borel_N}
\end{eqnarray}  
where the first term is similar to the collinear part in the event-shape case~(\ref{Borel_nu}).  Contrary to the latter, here the collinear singularity (appearing as a pole at $u=0$) requires a subtraction. This is the r\^ole of the second term in~(\ref{Borel_N}), which can be identified as the large-$N$ limit of the leading twist anomalous dimension. Having implemented the ${\overline{\rm MS}}$ factorization, $J$ is regular at $u=0$ and thus has a well defined perturbative expansion. Still, it has Borel singularities at $u=1,2$. 
These will be cancelled by the non-perturbative jet function $J_{\NP}(N\bar{\Lambda}^2/Q^2)$. Thus, the following ansatz suggests itself~\cite{DGE,DIS} 
\begin{eqnarray}
J_{\NP}\left({N\bar{\Lambda}^2}/{Q^2}\right) =
\exp\left\{-\omega_1\frac{C_F}{\beta_0}\frac{N\bar{\Lambda}^2}{Q^2}
-\frac12 \omega_2\frac{C_F}{\beta_0}\frac{N^2\bar{\Lambda}^4}{Q^4}\right\},
\label{C_NP}
\end{eqnarray}
where $\omega_i$ are non-perturbative parameters.

\section{Conclusions} 

Many interesting hard processes involve kinematic thresholds.
Owing to the emission of soft and collinear radiation, the corresponding differential cross-sections tend to have large perturbative and non-perturbative corrections.
As a result, a naive, fixed-order perturbative treatment is insufficient. Moreover, the OPE does not apply or tends to break down.  
Here we shortly reviewed the case of event-shape distributions in $e^+e^-$ annihilation, where we demonstrated the virtues of DGE both as a resummation method and as a way to study power corrections. This was followed by a deeper look into the case of DIS structure functions at large $x$, where the OPE was used to get additional insight into the problem.
There are many other physical applications where DGE and the shape-function approach can be applied, including, for example, fragmentation functions of light and heavy quarks and Drell--Yan or heavy-boson production processes. 

DGE is primarily a novel approach to resummation: the Sudakov exponent is calculated in a renormalization-scale-invariant manner by means of renormalon resummation. The criterion of a fixed-logarithmic accuracy becomes irrelevant when power corrections are being quantified: it is the subleading logs that carry the characteristic factorial growth of the coefficients, which is associated with the power corrections. Perturbatively, the additional resummation achieved by DGE with respect to the standard NLL resummation is significant. For event-shape distributions, it is~$\sim 20\%$ at $M_{\rm Z}$.
In principle, power corrections cannot be quantified without renormalon resummation. Our analysis of the thrust and the heavy-jet mass shows that this has very practical implications:  a consistent description of the two observables is possible only if renormalon resummation in the Sudakov exponent is performed (fig.~\ref{correlation}). Another important consequence is the significant impact on the extracted value of~$\alpha_s$.

DGE and the shape-function approach can be applied in the case of DIS structure function at large~$x$. However, here it can be contrasted with the OPE. As we have seen, 
a non-perturbative factorization can be consistent with the OPE, and it is supported by the OPE-based analysis.
In spite of the fact that the OPE tends to break down in the large-$x$ limit, it is very useful: a simple picture emerges from the analysis of twist~4 in terms of light-cone distributions. We have found that the dominant non-perturbative corrections at large $x$ are associated with the formation
of a narrow jet in the final state. These corrections are due to the 
exchange of gluons with momentum scale of the order~$W$, which are insensitive to the details of multiparton correlations in the target.
Instead of the full correlation, they measure a particular ingredient which is twist-2 like. It is therefore natural to conjecture that the dominant contributions at large~$x$ are associated with mixing with the leading twist. To $1/N$ accuracy the hadronization process of the jet involves a single target-dependent non-perturbative scale at each order in the twist expansion. These dominant corrections  can be resummed into a shape function of a single argument: $N/Q^2$, defining a non-perturbative jet function.  Thus, at large~$N$ the OPE collapses (to $1/N$ accuracy) into a factorized formula in which the leading twist is multiplied by a jet function~\footnote{It should be stressed that the factorization formula is 
not derived here from first principles. Its justification involves a strong assumption. Nevertheless, it is possible to check explicitly certain results. In particular, the formula predicts a common asymptotic behaviour of the logarithmic evolution in the large $N$ limit for any twist. This can be verified at twist~4.}. The application of DGE is then quite natural.

\section*{Acknowledgements}
It is a pleasure to thank my collaborators J.~Rathsman, G.P.~Korchemsky, D.A.~Ross and S.~Tafat for very enjoyable and fruitful collaboration. The research was supported in part by the EC program ``Training and Mobility of Researchers'', Network ``QCD and Particle Structure'', contract ERBFMRXCT980194.


\section*{References}
\begin{thebibliography}{99}

\bibitem{SVZ}
M.~A.~Shifman, A.~I.~Vainshtein and V.~I.~Zakharov,
%``QCD And Resonance Physics. Sum Rules,''
Nucl.\ Phys.\ B {\bf 147} (1979) 385.
%%CITATION = NUPHA,B147,385;%%

\bibitem{Beneke}
M.~Beneke,
%``Renormalons,''
{\em Phys. Rep.}  {\bf 317} (1999) 1 [hep-ph/9807443];
M.~Beneke and V.~M.~Braun,
%``Renormalons and power corrections,''
[hep-ph/0010208].

%dispersive approach
\bibitem{DMW}
Yu.~L.~Dokshitzer, G.~Marchesini and B.~R.~Webber,
%``Dispersive Approach to Power-Behaved Contributions in QCD Hard Processes,''
{\em Nucl. Phys.}  {\bf B469} (1996) 93
[hep-ph/9512336].
%%CITATION = HEP-PH 9512336;%%

\bibitem{Contopanagos:1996nh}
H.~Contopanagos, E.~Laenen and G.~Sterman,
%``Sudakov factorization and resummation,''
Nucl.\ Phys.\ B {\bf 484} (1997) 303
[hep-ph/9604313].
%%CITATION = HEP-PH 9604313;%%

\bibitem{CTTW}
S.~Catani, L.~Trentadue, G.~Turnock and B.~R.~Webber,
%``Resummation of large logarithms in e+ e- event shape distributions,''
{\em Nucl. Phys.}  {\bf B407} (1993) 3.
%%CITATION = NUPHA,B407,3;%%

\bibitem{Beneke:1995pq}
M.~Beneke and V.~M.~Braun,
%``Power corrections and renormalons in Drell-Yan production,''
{\em Nucl. Phys.}  {\bf B454} (1995) 253
[hep-ph/9506452].

\bibitem{Gardi:2001ny} 
E.~Gardi and J.~Rathsman,
%``Renormalon resummation and exponentiation of soft and collinear gluon  radiation in the thrust distribution,''
{\em Nucl. Phys.}  {\bf B609} (2001) 123
[hep-ph/0103217].
%%CITATION = HEP-PH 0103217;%%

\bibitem{DGE}
E.~Gardi,
%``Dressed gluon exponentiation,''
{\em Nucl. Phys.} {\bf B622} (2002) 365 
[hep-ph/0108222].
%%CITATION = HEP-PH 0108222;%%

\bibitem{Gardi:2002bg}
E.~Gardi and J.~Rathsman,
%``The thrust and heavy-jet mass distributions in the two-jet region,''
[hep-ph/0201019], to appear in {\em Nucl. Phys.} {\bf B}.
%%CITATION = HEP-PH 0201019;%%

\bibitem{CMW}
S.~Catani, B.~R.~Webber and G.~Marchesini,
%``QCD coherent branching and semiinclusive processes at large x,''
{\em Nucl. Phys.}  {\bf B349} (1991) 635.
%%CITATION = NUPHA,B349,635;%%

\bibitem{Korchemsky:1995is} 
G.~P.~Korchemsky and G.~Sterman,
%``Nonperturbative corrections in resummed cross sections,''
{\em Nucl. Phys.}  {\bf B437} (1995) 415
[hep-ph/9411211].
%%CITATION = HEP-PH 9411211;%%

\bibitem{Shape_function2} G.P. Korchemsky and G. Sterman,
Proc. 30th Rencontres de Moriond, {\em
QCD and high energy hadronic interactions}, Les Arcs,
France, 1995, ed. J. Tran Thanh Van (Editions
Fronti\`eres, Gif-sur-Yvette, 1995), p. 383 [hep-ph/9505391].

% shift of perturbative distribution
\bibitem{Dokshitzer:1997ew}
Yu.~L.~Dokshitzer and B.~R.~Webber,
%``Power corrections to event shape distributions,''
{\em Phys. Lett.}   {\bf B404} (1997) 321
[hep-ph/9704298].
%%CITATION = HEP-PH 9704298;%%

% shape function
\bibitem{Korchemsky:1999kt}
G.~P.~Korchemsky and G.~Sterman,
%``Power corrections to event shapes and factorization,''
{\em Nucl. Phys.}  {\bf B555} (1999) 335
[hep-ph/9902341].
%%CITATION = HEP-PH 9902341;%%

\bibitem{DIS}
E.~Gardi, G.~P.~Korchemsky, D.~A.~Ross and S.~Tafat,
%``Power corrections in deep inelastic structure functions at large  Bjorken x,'' 
[hep-ph/0203161], to appear in {\em Nucl. Phys.} {\bf B}.
%%CITATION = HEP-PH 0203161;%%

\bibitem{Dasgupta:2001sh}
M.~Dasgupta and G.~P.~Salam,
%``Resummation of non-global QCD observables,''
Phys.\ Lett.\ B {\bf 512} (2001) 323
[hep-ph/0104277].
%%CITATION = HEP-PH 0104277;%%

\bibitem{EVENT2} S.~Catani and M.H.~Seymour,
%``The Dipole Formalism for the Calculation of QCD Jet Cross Sections at Next-to-Leading Order,''
{\em Phys. Lett.}  {\bf B378} (1996) 287;
%%CITATION = HEP-PH 9602277;%%
%``A general algorithm for calculating jet cross sections in NLO QCD,''
{\em Nucl. Phys.} {\bf B485} (1997) 291;
%%CITATION = HEP-PH 9605323;%%
{\tt http://hepwww.rl.ac.uk/theory/seymour/nlo/ }

\bibitem{Korchemsky:2000kp} 
G.~P.~Korchemsky and S.~Tafat,
%``On power corrections to the event shape distributions in QCD,''
{\em JHEP} {\bf 0010} (2000) 010
[hep-ph/0007005].
%%CITATION = HEP-PH 0007005;%%

\bibitem{Belitsky:2001ij}
A.~V.~Belitsky, G.~P.~Korchemsky and G.~Sterman,
%``Energy flow in QCD and event shape functions,''
{\em Phys. Lett.}   {\bf B515} (2001) 297
[hep-ph/0106308].
%%CITATION = HEP-PH 0106308;%%


\bibitem{DGESHAPE}
E.~Gardi and J.~Rathsman, DGESHAPE,  A program for calculating the thrust and heavy-jet mass distributions using DGE, {\tt http://www3.tsl.uu.se/$\sim$rathsman/dgeshape/}.

\bibitem{Salam:2001bd}
G.~P.~Salam and D.~Wicke,
%``Hadron masses and power corrections to event shapes,''
{\em JHEP} {\bf 0105} (2001) 061
[hep-ph/0102343].
%%CITATION = HEP-PH 0102343;%%

%SDG
\bibitem{Average_thrust}
E.~Gardi and G.~Grunberg,
%``Power corrections in the single dressed gluon approximation:  The average thrust as a case study,''
{\em JHEP} {\bf 9911} (1999) 016
[hep-ph/9908458].
%%CITATION = HEP-PH 9908458;%%

\bibitem{Jaffe:zw}
R.~L.~Jaffe,
``Spin, twist and hadron structure in deep inelastic processes'',
[hep-ph/9602236].

\bibitem{JS}
R.~L.~Jaffe and M.~Soldate,
%``Twist Four In Electroproduction: Canonical Operators And Coefficient Functions,''
{\em Phys. Rev.}  {\bf D26} (1982) 49.

\bibitem{EFP}
R.~K.~Ellis, W.~Furmanski and R.~Petronzio,
%``Unravelling Higher Twists,''
{\em Nucl. Phys.}  {\bf B212} (1983) 29; and
%``Power Corrections To The Parton Model In QCD,''
% {\em Nucl. Phys.}  
{\bf B207} (1982) 1.

\bibitem{Jaffe}
R.~L.~Jaffe,
%``Parton Distribution Functions For Twist Four,''
{\em Nucl. Phys.}  {\bf B229} (1983) 205.

\bibitem{Balitsky:1989bk}
I.~I.~Balitsky and V.~M.~Braun,
%``Evolution Equations For QCD String Operators,''
{\em Nucl. Phys.}  {\bf B311} (1989) 541.

\bibitem{UnPub} V.~M.~Braun, unpublished notes.

\bibitem{Stein:1996wk}
E.~Stein, M.~Meyer-Hermann, L.~Mankiewicz and A.~Schafer,
%``IR-Renormalon Contribution to the Longitudinal Structure Function $F_L$,''
{\em Phys. Lett.}  {\bf B376} (1996) 177
[hep-ph/9601356].

\bibitem{DasW_DIS}
M.~Dasgupta and B.~R.~Webber,
%``Power Corrections and Renormalons in Deep Inelastic Structure Functions,''
{\em Phys. Lett.}  {\bf B382} (1996) 273 [hep-ph/9604388].

\bibitem{Maul:1997rz}
M.~Maul, E.~Stein, A.~Schafer and L.~Mankiewicz,
%``Phenomenology of IR-renormalons in inclusive processes,''
{\em Phys. Lett.}  {\bf B401} (1997) 100
[hep-ph/9612300].

\bibitem{UV_dom}
V.~M.~Braun,
``Ultraviolet dominance of power corrections in QCD?'',
[hep-ph/9708386].

\bibitem{BBM}
M.~Beneke, V.~M.~Braun and L.~Magnea,
%``Phenomenology of power corrections in fragmentation processes in e+ e- annihilation,''
{\em Nucl. Phys.}  {\bf B497} (1997) 297, [hep-ph/9701309].


\end{thebibliography}
\end{document}